\documentclass[sigconf]{acmart}

\usepackage{booktabs} 

\usepackage[linesnumbered,ruled,vlined]{algorithm2e}
\usepackage{amsmath}
\usepackage{enumitem}
\usepackage{amssymb,flushend}
\usepackage{bm}

\usepackage{color}

\usepackage{subcaption} 
\newcommand{\ra}[1]{\renewcommand{\arraystretch}{#1}}

\usepackage{tikz}
\usetikzlibrary{arrows,positioning,automata,calc,shapes}
\usepackage{pgfplots}
\pgfplotsset{compat=newest, scaled z ticks=false} 
\pgfplotsset{plot coordinates/math parser=false}
\newlength\figureheight 
\newlength\figurewidth

\usepackage[linesnumbered,ruled,vlined]{algorithm2e}

\acmDOI{10.475/123_4}

\acmISBN{123-4567-24-567/08/06}

\acmConference[KDD'19]{ACM Woodstock conference}{August 2019}{Anchorage, Alaska USA}
\acmYear{2019}
\copyrightyear{2019}

\acmArticle{4}
\acmPrice{15.00}

\begin{document}
\title{Is a Single Vector Enough?  Exploring Node Polysemy for Network Embedding}

\author{Ninghao Liu,$^1$ Qiaoyu Tan,$^1$ Yuening Li,$^1$ Hongxia Yang,$^2$ Jingren Zhou,$^2$ Xia Hu$^1$}
\affiliation{
 \institution{$^1$Department of Computer Science and Engineering, Texas A\&M University, TX, USA \\
 $^2$Alibaba Group, Hangzhou, China}
 }
\email{{nhliu43,qytan,liyuening,xiahu}@tamu.edu,  {yang.yhx, jingren.zhou}@alibaba-inc.com}

\begin{abstract}
Networks have been widely used as the data structure for abstracting real-world systems as well as organizing the relations among entities. Network embedding models are powerful tools in mapping nodes in a network into continuous vector-space representations in order to facilitate subsequent tasks such as classification and link prediction. Existing network embedding models comprehensively integrate all information of each node, such as links and attributes, towards a single embedding vector to represent the node's general role in the network. However, a real-world entity could be \textit{multifaceted}, where it connects to different neighborhoods due to different motives or self-characteristics that are not necessarily correlated. For example, in a movie recommender system, a user may love comedies or horror movies simultaneously, but it is not likely that these two types of movies are mutually close in the embedding space, nor the user embedding vector could be sufficiently close to them at the same time. In this paper, we propose a \textit{polysemous embedding} approach for modeling multiple facets of nodes, as motivated by the phenomenon of word polysemy in language modeling. Each facet of a node is mapped as an embedding vector, while we also maintain association degree between each pair of node and facet. The proposed method is adaptive to various existing embedding models, without significantly complicating the optimization process. We also discuss how to engage embedding vectors of different facets for inference tasks including classification and link prediction. Experiments on real-world datasets help comprehensively evaluate the performance of the proposed method.
\end{abstract}

%
%
%

\copyrightyear{2019} 
\acmYear{2019} 
\setcopyright{acmcopyright}
\acmConference[KDD '19]{The 25th ACM SIGKDD Conference on Knowledge Discovery and Data Mining}{August 4--8, 2019}{Anchorage, AK, USA}
\acmBooktitle{The 25th ACM SIGKDD Conference on Knowledge Discovery and Data Mining (KDD '19), August 4--8, 2019, Anchorage, AK, USA}
\acmPrice{15.00}
\acmDOI{10.1145/3292500.3330967}
\acmISBN{978-1-4503-6201-6/19/08}


\maketitle

\section{Introduction}\label{sec:intro}
Networks are ubiquitous data structures utilized for modeling information systems~\cite{Cui-etal17survey}, such as social networks, recommender systems, biological networks and knowledge graphs. In these systems, real-world entities such as users, items, molecules and knowledge concepts are abstracted as nodes in networks, while the relations between entities are modeled as links between them. The recent advances in network embedding propose to represent each node as a low-dimensional vector, by considering the node's neighborhood and feature information~\cite{Tang-etal15line, Tang-etal15PTE, Perozzi-etal14deepwalk, Grover-Leskovec17node2vec, Wang-etal16SDNE, Kipf-Welling17semisupervised}. Similar nodes are mapped close to each other in the embedding space. Node embedding has been shown to be an effective representation scheme that facilitates downstream network analysis tasks such as classification, clustering and link prediction~\cite{Cui-etal17survey, Hamilton-etal17representation}. 

\begin{figure}[t]
\centering
\includegraphics[scale=0.42]{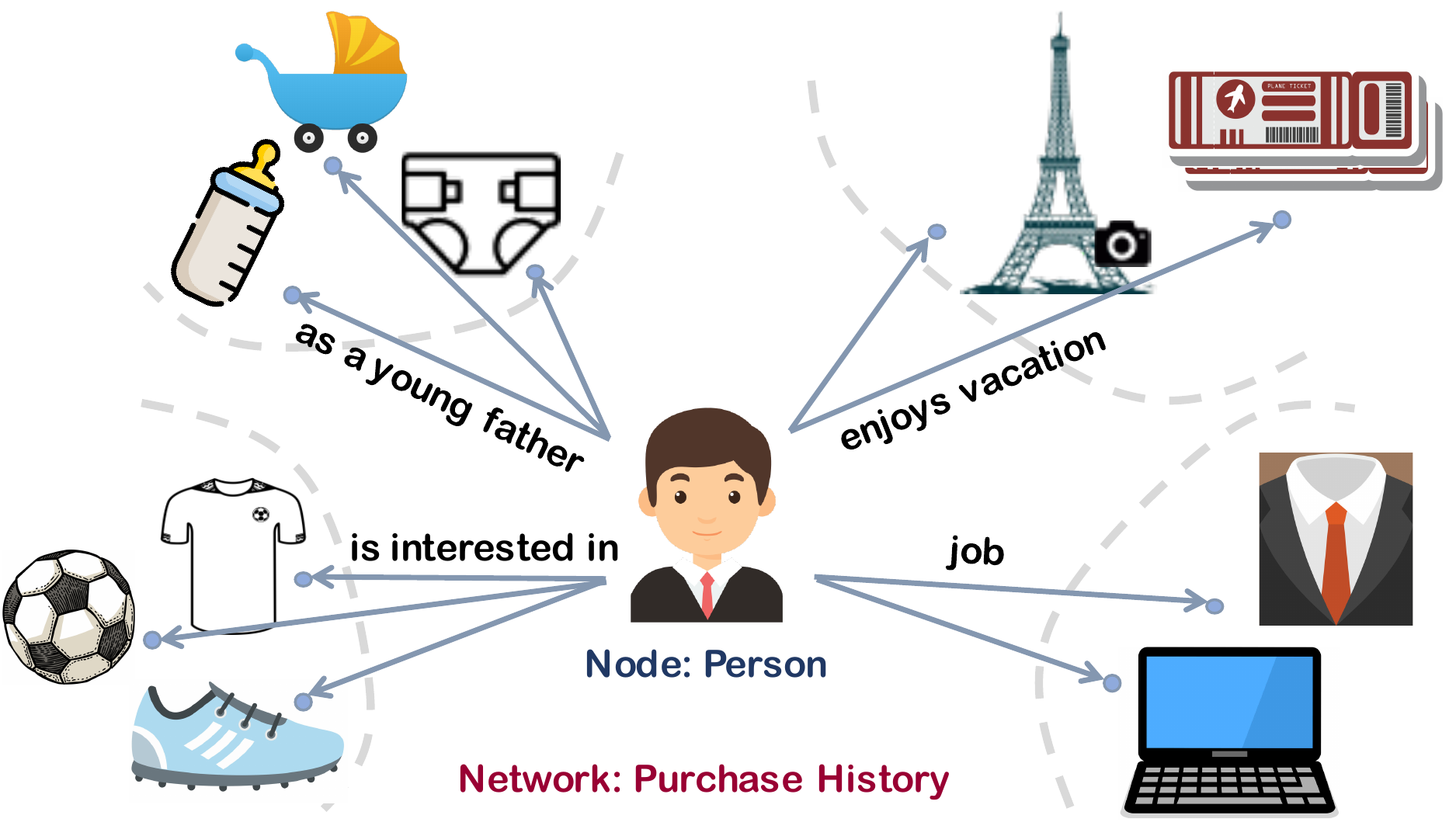}
\vspace{-3pt}
\caption{An example of multiple facets within a node. The network is built from online purchase history, where customers and items are nodes, and purchase actions are links.}
\label{fig:motiv}
\vspace{-10pt}
\end{figure}

It has been shown that, in many real-world applications, entities may possess disparate characteristics or aspects~\cite{perer2011visual, Reisinger_Mooney10multiPrototype}. In other words, a node in the network can be seen as a capsule containing different aspects. Different links stretching out from an entity to their neighbors could in fact result from the expression of its different aspects. In some scenarios, it could be problematic to fuse these disparate aspects into a single vector-space representation for a node. For example, in an online shopping website in Figure~\ref{fig:motiv} where customers and items are the nodes, a customer may have bought items of disparate genres. If we represent each customer with only a single vector, then the embedding vectors of the customer and items have to be simultaneously close to each other. It could be hard to achieve this, because other customers have different interests and it may mess up the distribution of embedding vectors. 

In this paper, we call such phenomenon as \textit{node polysemy}, where each node could have multiple facets, by making an analogy to a similar property possessed by words in natural language (e.g., ``bank'' can refer to either financial institutions or land near a river, depending on different contexts)~\cite{Reisinger_Mooney10multiPrototype, Huang-etal12multipleEmbedding}. In this setting, each node has multiple facets, while each facet of a node owns an embedding vector. In this work, we want to develop a polysemous network embedding method in order to discover multiple facets of nodes and learn representations for them.

The challenges of developing polysemous network embedding models are three-fold. The first is how to determine the facets of nodes, as well as to flexibly update embeddings of different facets in the vector space. For each data sample (e.g., links or random walks), we need to determine which facet of each node is likely to be activated, so as to update the corresponding embedding of that facet in the training process. The second challenge is how to maintain the correlation among embedding vectors of different facets. Although we split a node into multiple facets, different facets may not be completely uncorrelated to each other. Some information will be lost if we simply model each facet independently. Third, when considering node polysemy, how to make the modeling process adaptive to the existing well-established base models such as  Deepwalk~\cite{Perozzi-etal14deepwalk}, LINE~\cite{Tang-etal15line}, PTE~\cite{Tang-etal15PTE} and GCN~\cite{Kipf-Welling17semisupervised}. Besides, the computational complexity of the new model will inevitably increase when different facets of nodes are considered. Therefore, an efficient optimization algorithm is also needed, especially when considering its compatibility to negative sampling. 

Specifically, in this paper, we propose a polysemous network embedding method in order to take into account multiple facets of nodes in network data. Each facet of a node is represented using an embedding vector. The correlation between different node facets can be preserved in our method. The developed modeling strategy is flexible to be applied to various base embedding models without making radical changes to their base formulation. We first show how to revise Deepwalk to tackle node polysemy, and then extend our discussion to more base embedding models and more complex application scenarios. The optimization process for training the polysemous embedding models is specially designed to guarantee its efficiency and implementation feasibility, especially when applying negative sampling. Finally, to evaluate whether considering multiple aspects of nodes could actually benefit learning node representations and downstream data mining tasks, we conduct experiments with different tasks to compare the performance between polysemous embedding models and their corresponding single-vector base models. The contribution of this paper is summarized as follows:
\begin{itemize}
\item We propose a novel polysemous network embedding method to incorporate different facets of nodes into representation learning. Each aspect of a node owns an embedding vector, and the relations among embedding vectors of different aspects are also considered.
\item We specifically formulate the problem to guarantee that the resultant optimization algorithm is efficient and feasible for implementation. Also the proposed method is flexible to be adapted to various existing transductive embedding models.
\item We conduct intensive experiments considering different downstream tasks and application scenarios, providing insights on how and when we benefit from modeling node polysemy in the network embedding.
\end{itemize}

 
\section{Polysemous Network Embedding}\label{sec:link}
In this section, we introduce the core idea of polysemous network embedding, by using Deepwalk as the base model. Then, we design an optimization algorithm to train the polysemous embedding model. We will also discuss how to estimate the facet of a node in different contexts. Finally, we introduce how to combine embedding vectors of different facets towards downstream tasks such as classification and link prediction. 

\subsection{Polysemous Deepwalk}\label{sec:polysg}
In the setting of polysemous embedding, for the Deepwalk model, each node $v_i$ is associated with a target embedding matrix $\textbf{U}_i \in \mathbb{R}^{K_i \times D}$ and a context embedding matrix $\textbf{H}_i \in \mathbb{R}^{K_i \times D}$. Here $K_i$ is the number of embedding vectors possessed by node $v_i$ considering its different possible facets. The traditional Deepwalk model could be seen as a special case where $K_i = 1$. The embedding vector for facet $k$ of node $v_i$ is denoted as $\textbf{U}^k_i \in \mathbb{R}^{D}$ or $\textbf{H}^k_i \in \mathbb{R}^{D}$, and $D$ is the dimension of each embedding vector. Different nodes may be associated with different number of embedding vectors, depending on the diversity of their characteristics in the network. In this work, for illustration purposes, we simply let all nodes have the same number of embedding vectors, so that $K_i = K$ and $K$ is a predefined constant integer. In practice, the value of $K$ could be estimated from data. For examples, $K$ can be approximately set as the number of latent interest categories in recommender systems, or be estimated as the number of major topics in an academic network. We will discuss in later sections about how to assign facets to nodes with different probabilities.

Deepwalk utilizes the skip-gram model~\cite{Mikolov-etal12word2vec} that is trained using the maximum likelihood estimation, where we try to find the model parameters that maximize the likelihood of obtained \textit{observations}. Specifically, let $\theta$ be the parameters to be optimized and $\mathcal{O}$ be the set of all observations, the objective function to be maximized is:
\begin{equation}\label{eq:overall}
\begin{split}
\mathcal{L}_{DW}(\theta) &= \sum_{o \in \mathcal{O}} \log\ p(o |\theta) = \sum_{o \in \mathcal{O}} \log\ p\big( ( \mathcal{N}(v_i), v_i ) | \theta\big) \\
& = \sum_{o \in \mathcal{O}} \sum_{v_j\in \mathcal{N}(v_i)} \log\ p(v_j | v_i) ,
\end{split}
\end{equation}
where each observation $o \in \mathcal{O}$ is defined as a tuple, $o=( \mathcal{N}(v_i), v_i )$, consisting of a central node $v_i$ and its context $\mathcal{N}(v_i)$. Each node within the context is denoted as $v_j$ so that $v_j\in \mathcal{N}(v_i)$. The model parameters, i.e., the embedding vectors of nodes, are used in computing $p(v_j | v_i)$ which is the conditional probability of having $v_j$ as one of the contexts given $v_i$.

\begin{figure*}[t]
\centering
\includegraphics[scale=0.40]{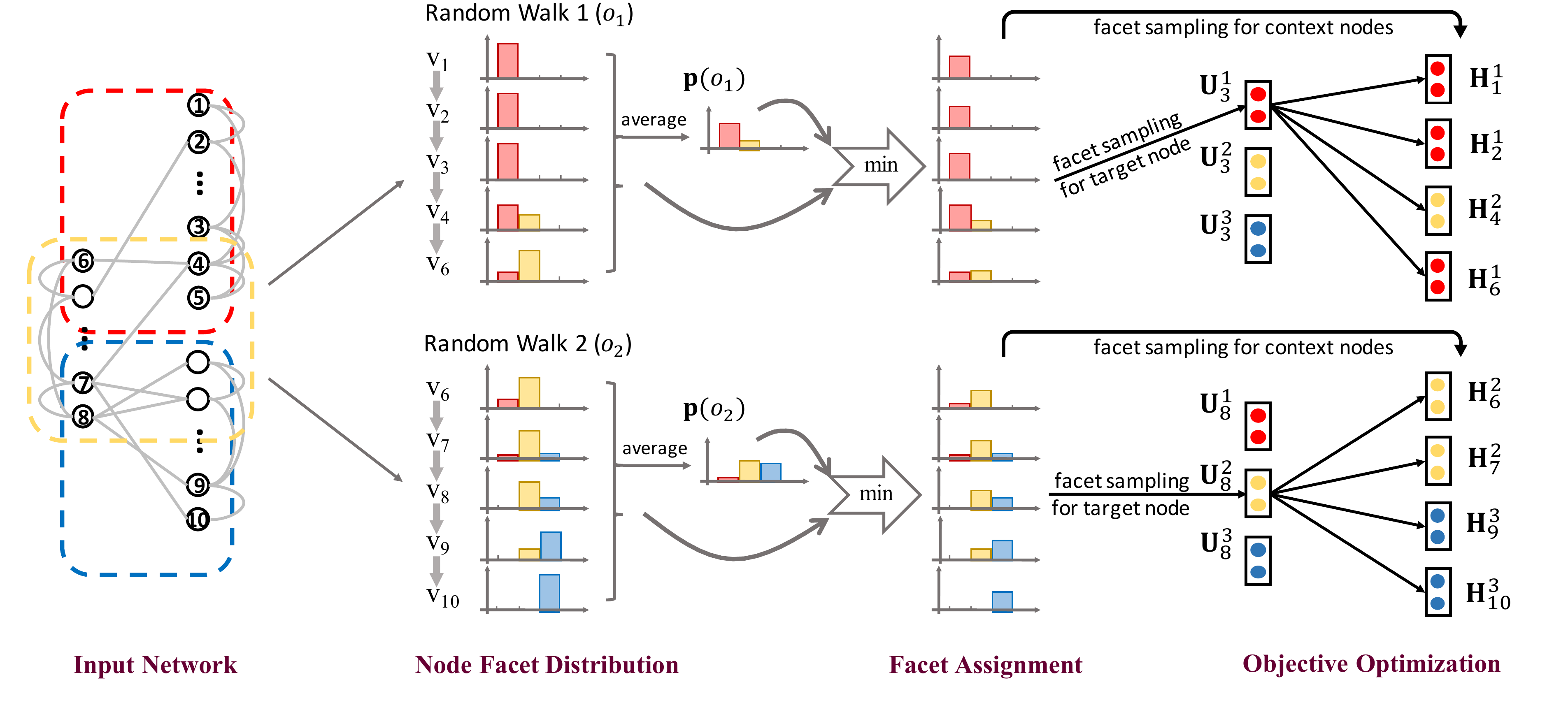}
\vspace{-10pt}
\caption{The overall training procedure of Polysemous Deepwalk. Each color refers to one facet contained in the input network. Each histogram visualizes a probability distribution. Note that the above network is only a toy example for illustration purposes, where numerical values do not reflect real results.}
\label{fig:overall}
\vspace{-5pt}
\end{figure*}

However, in our settings where each node owns multiple facets, the activated facet of a node varies under different contexts. Also, the facet of a given context is determined by the combination of all facets of the nodes within the context. Suppose the distribution of node facets are known in advance, and we treat them as prior knowledge denoted as $\mathcal{P}$. Taking the additional information into account, the objective is reformulated as:
\begin{equation}\label{eq:overall_new}
\begin{split}
\mathcal{L}_{PolyDW}(\theta) &= \sum_{o \in \mathcal{O}} \log\ p(o |\mathcal{P}, \theta) \\
& = \sum_{o \in \mathcal{O}} \log\ [\sum_{s(o)} p(o |s(o), \mathcal{P}, \theta) \cdot p(s(o) | \mathcal{P}, \theta) ] .
\end{split}
\end{equation}
Here $s(o)$ is defined as one case of activated facets of all nodes within $o$, so $s(o) = \{ s(v|o)\ | v\in v_i \cup \mathcal{N}(v_i) \}$ where $s(v|o)$ is the activated facet of node $v$ in the context of $o$. In a given observation $o$, suppose the activated facet of $v_i$ is $k_i$ and the facet of each $v_j \in \mathcal{N}(v_i)$ is $k_j$, the conditional probability $p(o |s(o), \mathcal{P}, \theta)$ is thus defined as:
\begin{equation}\label{eq:observation}
\begin{split}
p(o |s(o), \mathcal{P}, \theta) = \prod_{v_j\in \mathcal{N}(v_i)} p(v_j | v_i, s(o)) ,
\end{split}
\end{equation}
and each product factor is calculated as
\begin{equation}\label{eq:cond}
p(v_j | v_i, s(o)) = \frac{\exp(<\textbf{H}^{k_j}_j, \textbf{U}^{k_i}_i>)}{\sum_{v, k} \exp(<\textbf{H}^{k}_v, \textbf{U}^{k_i}_i>)},
\end{equation}
which is similar to the softmax function in traditional skip-gram models, except that general node embeddings are replaced by node facet embeddings. Here "$< \cdot \, , \cdot  >$" denotes the inner product of two vectors. The denominator acts as normalization over all possible nodes and facets. For readability, we omit $\mathcal{P}$ after the expansion in Equation~\ref{eq:observation}, as it is no longer applied in later steps.

\setlength{\textfloatsep}{3pt}
\begin{algorithm}[t]
\small
	\DontPrintSemicolon
	\SetKwFunction{algo}{algo}\SetKwFunction{proc}{proc}
	\KwIn{Input network $\mathcal{G}$, number of nodes $N$, facet sampling rate $R$.}  
	\KwOut{Embedding matrix $\textbf{U}$, context embedding matrix $\textbf{H}$.}
	\SetKwProg{myalg}{Algorithm}{}{}
	    Initialize $\textbf{U}$ and $\textbf{H}$ ;\\
	    Estimate facet distribution $\textbf{p}(v_i)$ for each node, so that $\mathcal{P} = \{\textbf{p}(v_i)\}$, $1\le i \le N$ ;\\
	    Obtain observations $\mathcal{O}$ via random walks and context window sliding ;\\
		\While{$\mathrm{not \: converged}$}{
			\For{$o \in \mathcal{O}$}{
			    Obtain a target-context tuple $o = (\mathcal{N}(v_i), v_i)$ ;\\
				\For{$1 \le r \le R$}{
				    Sample a facet $k_i = s(v_i|o) \sim \textbf{p}(v_i|o)$, defined in Eq.~\ref{eq:minpp} ;\\
				    Sample a facet $k_j = s(v_j|o) \sim \textbf{p}(v_j|o)$, defined in Eq.~\ref{eq:minpp}, for each context node $v_j\in \mathcal{N}(v_i)$ ;\\
				    Obtain negative samples ;\\
				    Update $\textbf{U}^{k_i}_i$, $\textbf{H}^{k_j}_j$ for $v_j\in \mathcal{N}(v_i)$, as well as facet embeddings of negative samples, using stochastic gradient descent ;\
				}
				
			}
		}
	\caption{Polysemous Deepwalk}
	\label{alg:polydeepwalk}
\end{algorithm}

It is cumbersome to directly apply gradient descent to optimize the objective function in Equation~\ref{eq:overall_new} $\sim$ Equation~\ref{eq:cond}, due to the summation term inside the logarithm function. Also, it becomes unclear of how to incorporate negative sampling~\cite{Mikolov-etal12word2vec} to approximate the normalization term for more efficient computation. Therefore, we further derive the objective function as below:
\begin{equation}\label{eq:overall_final}
\begin{split}
\mathcal{L}_{PolyDW}(\theta) &= \sum_{o \in \mathcal{O}} \log\ [\sum_{s(o)} p(o |s(o), \mathcal{P}, \theta) \cdot p(s(o) | \mathcal{P}, \theta) ] \\
&\geq \sum_{o \in \mathcal{O}} \sum_{s(o)} p(s(o) | \mathcal{P}, \theta) \cdot \log\ p(o |s(o), \mathcal{P}, \theta) \\
&= \sum_{o \in \mathcal{O}} \sum_{s(o)} p(s(o) | \mathcal{P}) \cdot [\sum_{v_j\in \mathcal{N}(v_i)} \log\ p(v_j | v_i, s(o))] \\
&= \mathcal{L}^*_{PolyDW}(\theta)
\end{split}
\end{equation}
The intuition behind the above transformation is that, instead of maximizing the original objective function, we propose to maximize its lower bound~\cite{Kingma-Welling15autoVarBayes}, denoted as $\mathcal{L}^*_{PolyDW}(\theta)$, by applying Jensen's inequality. The final form of the objective function is similar to that of the skip-gram model, except the external summation over $s(o)$. As a result, we could adopt the same negative sampling strategy as in the traditional skip-gram model to approximate the normalization term in $p(v_j | v_i, s(o))$. Therefore, one major advantage of the proposed polysemous embedding model is that the training process can be easily implemented through minimal modification to the existing learning frameworks, by adding an additional sampling step of assigning activated facets to nodes in each observation $o$.

Specifically, the summation over $\sum_{s(o)}$, following the distribution of $p(s(o)| \mathcal{P})$, is implemented through \textit{facet sampling} separately for each node in $o$. The overall optimization algorithm is specified in Algorithm~\ref{alg:polydeepwalk}. After initialization and node-facet distribution estimation (will be introduced later), a number of random walks are sampled just as in traditional Deepwalk (Line 3). Then, for each observation $o$, several rounds of facet sampling are conducted on each node within $o$ (the loop in Line $7$). In each round, each node has one facet activated (Line $8\sim 10$), so that the embedding vector corresponding to that facet will be updated using SGD (Line 11). The major additional computation cost, compared with traditional Deepwalk, comes from the increased size of training data in $\mathcal{O}$ by a factor of sampling rate $R$.

The overall process of polysemous Deepwalk is illustrated in Figure~\ref{fig:overall}, where ``Objective Optimization" has been introduced as above, while the other steps related with facet distribution and facet assignment will be discussed in detail in the next subsection.

\subsection{Node-Facet Assignment} \label{sec:nodefacet}
We now introduce how the prior knowledge $\mathcal{P}$ can be obtained, and how to determine the facet of nodes given a specific observation $o$.
For now, we limit the discussion to undirected homogeneous plain networks, and provide one method to obtain the global distribution of node facets by only leveraging the network adjacency matrix. For networks with attribute information or heterogeneous information networks, we may resort to other strategies and we leave it to future work. Let $\textbf{A}$ be the symmetric adjacency matrix of the network, we perform community discovery on the network~\cite{Kuang-etal12symmetric}\cite{WXu-etal01document}:
\begin{equation}
\min_{\textbf{P}\ge 0}\ \| \textbf{A} - \textbf{P} \cdot \textbf{P}^T \|^2_F + \alpha \| \textbf{P} \|^2_F,
\end{equation}
where $\textbf{P}$ can be solved using gradient search algorithms. The probability $p(k|v)$, that node $v$ is assigned with the $k$-th facet, can be calculated as $p(k|v) = \frac{\textbf{P}_{v,k}}{\sum_{c=1,...,K} \textbf{P}_{v,c}}$. We define the facet distribution of a node as $\textbf{p}(v)=[p(1|v),...,p(K|v)]$. We treat $\textbf{p}(v_i), 1 \le i \le N$ as the prior knowledge $\mathcal{P}$, as it encodes the global understanding of the network states. Applying attribute information could achieve better estimation, but it is beyond the discussion in this work.

After obtaining the prior knowledge $\mathcal{P}$, we are able to estimate the overall facet of each observation $o$ as well as the facet of nodes within $o$. Given an observation $o=(\mathcal{N}(v_i), v_i)$, a straightforward way to define its facet distribution $\textbf{p}(o)$ is by averaging the facet distributions of nodes inside the observation, i.e., $\textbf{p}(o) = \big( \textbf{p}(v_i) + \sum_{v_j\in \mathcal{N}(v_i)} \textbf{p}(v_j) \big)/(|\mathcal{N}(v_i)|+1)$. Considering that the activated facet of a node depends on the specific context where it is observed, given $o$, the node's facet $s(v|o)$ is sampled according to the distribution $\textbf{p}(v|o)$, which is defined heuristically as below
\begin{equation}\label{eq:minpp}
    \textbf{p}(v|o) = \min (\textbf{p}(v), \textbf{p}(o)),
\end{equation}
where we include a $\min(\cdot,\cdot)$ operator because it is undesirable to assign a node $v_i$ with facet $k$ if $p(k|v_i)\approx 0$, even when the $k$-th entry in $\textbf{p}(o)$ is large. To make it a valid probability distribution, $\textbf{p}(v|o)$ is further normalized to sum to 1. 

Till now, we have introduced the whole training process of polysemous Deepwalk as shown in Figure~\ref{fig:overall}. Given the input network, we first estimate node facet distributions as prior knowledge $\mathcal{P}$. Then, random walks are performed to construct node-context observations $\mathcal{O}$. After that, within each walk sample $o$, each node is assigned with an activated facet. Finally, node embeddings of correspondent facets are updated through optimization.

\subsection{Joint Engagement of Multiple Embeddings for Inference}\label{subsec:joint} We are able to obtain multiple embedding vectors for each node after training the polysemous model. Then the question arises that, during inference, how to collectively consider different embedding vectors for subsequent tasks. Here we discuss two major network analysis tasks including classification and link prediction.

For classification, for each node, our strategy is to combine multiple vectors $\{\textbf{U}^k_i\}_{k=1,...,K}$ into a joint vector $\Tilde{\textbf{U}_i}$. Specifically, some options include:
\begin{equation}
    \Tilde{\textbf{U}}_i = \textbf{U}^1_i \oplus \textbf{U}^2_i \oplus ... \oplus \textbf{U}^K_i ,
\end{equation}
where we directly concatenate embeddings of different facets, or
\begin{equation}\label{eq:weightconcat}
    \Tilde{\textbf{U}}_i = \big( p(1|v_i) \cdot \textbf{U}^1_i \big) \oplus \big( p(2|v_i) \cdot \textbf{U}^2_i \big) \oplus ... \oplus \big( p(K|v_i) \cdot \textbf{U}^K_i \big) ,
\end{equation}
where we first scale each embedding vector with the probability of belonging to the corresponding facet and then concatenate these scaled vectors. Here $\oplus$ denotes the concatenation operation. The resultant embedding vectors $\{\Tilde{\textbf{U}}_i\}_{i=1,...,N}$ can be directly used in node classification. We adopt Equation~\ref{eq:weightconcat} in experiments.

For link prediction or network reconstruction tasks, a higher similarity score between the representations of two nodes indicates greater possibility that a link exist between the nodes. We can define the similarity between two nodes $v_i$ and $v_j$ as:
\begin{equation}
similarity(v_i, v_j) = \sum^K_{k=1} \sum^K_{k'=1} p(k|v_i)\ p(k'|v_j) <\textbf{U}^k_i, \textbf{U}^{k'}_j> ,
\end{equation}
where different facet pairs of embedding vectors contribute to the overall similarity computation, weighted by the probability that the node belongs to the corresponding facet.

\subsection{Discussion}
The proposed work can be regarded as an example of disentangled representation learning~\cite{higgins2017beta}, since different representation dimensions are sensitive to varying factors behind the data. Moreover, it helps improving the interpretability of representation learning~\cite{du2018techniques}, because representation dimensions are separated according to node facets that could be associated with concrete meanings or characteristics of real-world networked objects.

\section{Models Extended by Polysemous Embedding}
The methodology of polysemous embedding can also be applied to extend other fundamental single-embedding models. In this section, we elaborate two scenarios as examples. First, we show how polysemous embedding can be used in heterogeneous networks where links exist between nodes of different types. Second, we show how polysemous embedding can be combined with graph neural networks, where the embedding look-up tables in Deepwalk is replaced by feedforward modules.

\subsection{Polysemous PTE for Heterogeneous Networks}\label{sec:polypte}
We show how to consider node polysemy when modeling heterogeneous networks. We choose PTE~\cite{Tang-etal15PTE} as the base model to be extended, as it is a fundamental model for tackling the problem. To simplify the illustration, we only consider bipartite networks during model development. The discussion can be applied for more complex networks, since the objective function of PTE could be extended as the sum of several terms considering multiple bipartite or homogeneous networks.

Given a network containing two types of nodes $\mathcal{V}_A$ and $\mathcal{V}_B$, as well as a set of links denoted as $\mathcal{E}$, each observation $o$ is defined as a link $o=(u_j, v_i)\in \mathcal{E}$ where $v_i\in \mathcal{V}_A, u_j\in \mathcal{V}_B$. The set of all observations is thus defined as $\mathcal{O} = \mathcal{E}$. Similar to the derivation in Section~\ref{sec:polysg}, the objective function is formulated as:
\begin{equation}\label{eq:pte}
\begin{split}
\mathcal{L}_{polyPTE}(\theta)
&= \sum_{o \in \mathcal{E}} \log\ [ \sum_{s(o)} p(o|s(o), \mathcal{P}, \theta) ) \cdot p(s(o) | \mathcal{P}, \theta)] \\
&\geq \sum_{o \in \mathcal{E}} \sum_{s(o)} p(s(o) | \mathcal{P}, \theta) \cdot \log\ p(o|s(o), \mathcal{P}, \theta) \\
&= \sum_{o \in \mathcal{E}} \sum_{s(o)} p(s(o) | \mathcal{P}) \cdot \log\ p(u_j | v_i, s(o)),
\end{split}
\end{equation}
where 
\begin{equation}
p(u_j | v_i, s(o)) = \frac{\exp(<\textbf{H}^{k_j}_j, \textbf{U}^{k_i}_i>)}{\sum_{v, k} \exp(<\textbf{H}^{k}_v, \textbf{U}^{k_i}_i>)} .
\end{equation}
According to the original PTE model, the matrix $\textbf{U}$ contains the embedding vectors of nodes in $\mathcal{V}_A$, and $\textbf{H}$ contains the embedding vectors of nodes in $\mathcal{V}_B$. The resultant lower bound objective is denoted as $\mathcal{L}^*_{polyPTE}$, which is to be optimized.

\subsubsection{Node Facet Assignment} 
The strategy of determining heterogeneous node facet distribution is similar to that of the previous section. Given the asymmetric adjacency matrix $\textbf{A}$, we first solve
\begin{equation}
\min_{\textbf{P}\ge 0, \textbf{Q}\ge 0} \| \textbf{A} - \textbf{P} \cdot \textbf{Q}^T \|^2_F + \alpha (\| \textbf{P} \|^2_F + \| \textbf{Q} \|^2_F) ,
\end{equation}
where we set $\alpha = 0.05$ in experiments. The resultant factor matrix $\textbf{P}$ contains the association intensity between $\mathcal{V}_A$ and facets, while $\textbf{Q}$ contains the facet association information for nodes in $\mathcal{V}_B$. Then, we normalize $\textbf{P}$ and $\textbf{Q}$ to obtain the probabilities of belonging to different facets for each node $v_i\in \mathcal{V}_A$ and $u_j\in \mathcal{V}_B$~\cite{WXu-etal01document}. If we denote the facet distribution as $\textbf{p}(v_i)$ and $\textbf{p}(u_j)$ respectively for the two types of nodes, then the facet distribution of an observation (i.e., an edge) is computed as $\textbf{p}(o) = \big(\textbf{p}(v_i) + \textbf{p}(u_j)\big)/2$. Different from polysemous Deepwalk, here we simply let $\textbf{p}(v|o) = \textbf{p}(o)$ be applied for node-facet sampling, because the window size in PTE equals to $1$ which is much smaller than Deepwalk.

\subsubsection{Engage Multiple Embeddings for Inference} The way of jointly considering a node's multiple embedding vectors, for downstream tasks, is similar to what is introduced in Section~\ref{subsec:joint}. The only difference is that the measurement of similarity between two different types of nodes is adjusted as:
\begin{equation}
similarity(v_i, u_j) = \sum^K_{k=1} \sum^K_{k'=1} p(k|v_i)\ p(k'|u_j) <\textbf{U}^k_i, \textbf{H}^{k'}_j> .
\end{equation}

\subsection{Polysemous Embedding with GCN}
The idea of polysemous embedding can be realized with models of other architectures. In this subsection, we consider GCN~\cite{Kipf-Welling17semisupervised} as an example and show how it could be extended as PolyGCN to consider node polysemy. Different from PolyDeepwalk and PolyPTE that keep embedding look-up tables to store embedding vectors, PolyGCN uses a feedforward network module to generate embedding of each node by gathering information from neighborhoods.

The core step of embedding generation in GCN is to aggregate information from a node's local neighborhood. Let $\textbf{u}_d(i)$ denote the intermediate representation of node $v_i$ on layer of depth $d$, then the forward propagation is
\begin{equation}
    \textbf{u}_d(i) \leftarrow \sigma \bigg(\textbf{W}_d \cdot \text{MEAN}(\textbf{u}_{d-1}(i) \cup \{\textbf{u}_{d-1}(j), \forall v_j\in\mathcal{N}(v_i)\}) \bigg) , \,\,\,\,\,\,\,
\end{equation}
where $\textbf{W}_d$ is the weight matrix on layer $d$, MEAN refers to the aggregator based on the pre-processed adjacency matrix, and $\mathcal{N}(v_i)$ is the local neighborhood of $v_i$. The final embedding output for $v_i$ is defined as $\textbf{U}_i = \textbf{u}_{d_{max}}(i)$, where $d_{max}$ denotes the depth of the final layer. After forward propagation, GCN can be trained in an unsupervised manner similar to Deepwalk or PTE.

We here propose a feasible approach to incorporate node polysemy into GCN-based models, where each facet of embeddings corresponds to one GCN model, while the internal architecture of each GCN is unchanged. Specifically, within a certain facet $k$, we limit the local neighborhood of the node as other nodes with the same facet $k$ activated, so that
\begin{equation}
    \textbf{u}^k_d(i) \leftarrow \sigma \bigg(\textbf{W}^k_d \cdot \text{MEAN}(\textbf{u}^k_{d-1}(i) \cup \{\textbf{u}^k_{d-1}(j), \forall v_j\in\mathcal{N}^k(v_i)\}) \bigg) , \,\,\,\,\,\,\,
\end{equation}
where $\mathcal{N}^k(v_i)$ denotes the local neighborhood under facet $k$. The final embedding output for facet $k$ of $v_i$ is $\textbf{U}^k_i = \textbf{u}^k_{d_{max}}(i)$. The main question to be answered is how to construct the local neighborhood for a specific facet $k$ of node $v$. Here we still consider bipartite networks as the scenario, so for each facet there are two GCNs respectively for generating embeddings for each type of nodes. Following the similar strategy as in Section~\ref{sec:polypte}, the observed links are decomposed into interaction results of different facet pairs. We further relax the problem so that embeddings of different facets are mutually independent. Specifically, the original observation matrix $\textbf{A}=\sum_k \textbf{A}^k$, where $\textbf{A}^k \ge 0$ and $\textbf{A}^k(i,j)$ is proportional to $\textbf{P}(i,k)\cdot \textbf{Q}(j,k)$. $v_j\in\mathcal{N}^k(v_i)$ if both $v_i$ and $v_j$ connect to the same node of the other type under facet $k$. For training each GCN pair, we adopt the similar unsupervised training strategies as proposed in~\cite{Hamilton-etal17inductive}, where nodes are encouraged to have similar embeddings of facet $k$ if they are connected according to the corresponding observation matrix.



\section{Experiments}\label{sec:exp}
We try to answer several questions through experiments. First, how effective is the proposed polysemous embedding method compared with single-vector embedding counterparts? Second, is considering node polysemy beneficial for network embedding when evaluated on different downstream tasks? Third, how will polysemous embedding models react to the changes of hyperparameters?
\subsection{Experimental Settings}
We first introduce the applied datasets as below. The detailed information is shown in Table~\ref{table:datasets}.
\begin{itemize}[leftmargin=*]
    \item \textbf{BlogCatalog}: A social network dataset built based on connections between online bloggers. The original dataset contains both link and attribute information, while we only keep the links in our experiments. Each node is also associated with a label determined from some predefined interest categories of bloggers.
    \item \textbf{Flickr}: A network dataset constructed from interactions between users on a multimedia hosting website. Each link corresponds to a following relation between users. The groups that users joined are regarded as labels.
    \item \textbf{MovieLens}: A movie rating dataset widely used for evaluating collaborative filtering models. In our experiments, we treat it as a heterogeneous network where links exist between users and items (i.e., movies). We transform rating scores into implicit data, so that each entry is either $0$ or $1$ indicating whether the user rated the movie~\cite{he2016fast}. In this way, conducting recommendation on this dataset can also be regarded as performing link prediction. 
    \item \textbf{Pinterest}: An implicit feedback data originally constructed for image recommendation. Users and images are regarded as nodes. Each link is a pin on an image initiated by a user. Each user has at least $20$ links. Similarly, the link prediction task can be seen as recommending images to users.
\end{itemize}
The networks in BlogCatalog and Flickr are homogeneous, and we use them in classification and link prediction. The networks in MovieLens and Pinterest are heterogeneous, and we will apply them in link prediction tasks which can also be seen as doing recommendation. The baseline methods for comparison are as below.
\begin{itemize}[leftmargin=*]
    \item \textbf{Deepwalk}~\cite{Perozzi-etal14deepwalk}, \textbf{PTE}~\cite{Tang-etal15PTE}, \textbf{GCN}~\cite{Ying-etal18graph}: Some commonly used network embedding models that map each node to a single vector. We include them as baseline methods to analyze whether their polysemous embedding counterparts achieve better performance. Here GCN is only applied in heterogeneous link prediction.
    \item \textbf{NMF}~\cite{Lee-Seung01nmf, Kuang-etal12symmetric}: A traditional model that learns latent factors from data. We include NMF as one of the baseline methods because we applied it in estimating the global facets contained in data. The number of latent factors is the same as the number of facets in the proposed polysemous model.
    \item \textbf{MSSG}~\cite{Neel-etal15efficient}: A multi-sense word embedding model original developed for natural language processing. Senses of words determined by clustering their average context representations. Each word has only one embedding vector when it is regarded as context. The model is adapted for network analysis tasks.
    \item \textbf{MNE}~\cite{Yang-etal18multifacet}: A model that factorizes the network proximity matrix into several group of embedding matrices, and adds a diversity constraint to force different matrices focus on different aspects of nodes. We only use MNE for node classification since the way of performing link prediction is not discussed in the paper.
\end{itemize}

\begin{table}[t!]
\captionsetup{font=small}
\small
	\centering
      \begin{tabular}[0.1\textwidth]{c|c|c|c}
      \hline \hline
      \multicolumn{1}{c}{\bfseries $\,\,\,\,$} & \multicolumn{1}{|c}{\bfseries $\,\,\,\,|\mathcal{V}_A|\,\,\,\,$ } & \multicolumn{1}{|c}{\bfseries $\,\,\,\,|\mathcal{V}_B|\,\,\,\,$ } & \multicolumn{1}{|c}{\bfseries $\,\,\,\,|\mathcal{E}|\,\,\,\,$}\\
      \hline 
      \textbf{ BlogCatalog }  & $5$,$196$ & $-$ & $171$,$743$ \\
      \hline
      \textbf{ Flickr } & $7$,$575$ & $-$ & $239$,$738$ \\
      \hline
      \textbf{ MovieLens } & $6$,$040$ & $3$,$952$ & $1$,$000$,$209$ \\
      \hline
      \textbf{ Pinterest } & $55$,$187$ & $9$,$916$ & $1$,$500$,$809$ \\
      \hline  \hline
      \end{tabular}
     \vspace{-0pt}
	\caption{Statistics of datasets.} \label{table:datasets}
\vspace{-5pt}
\end{table}

\subsection{Node Classification}
Node classification is a commonly applied task for evaluating network embedding outcomes. In this experiment, we compare the performance of the proposed polysemous models compared with baseline models. We use Deepwalk as the base single-embedding model. The proposed model is named as PolyDeepwalk. The BlogCatalog and Flickr datasets are applied in this experiment. 

\begin{figure}[t]
\captionsetup{font=small}
\captionsetup[subfigure]{justification=centering}
\hspace{0pt}
\begin{subfigure}[b]{0.24\textwidth}
  \setlength\figureheight{1.00in}
  \setlength\figurewidth{1.35in}
  \centering  \scriptsize
%
%
\definecolor{mycolor1}{rgb}{0.00000,0.44700,0.64100}%
\definecolor{mycolor2}{rgb}{0.65000,0.22500,0.09800}%
\definecolor{mycolor3}{rgb}{0.14700,0.59800,0.17500}%
\definecolor{mycolor4}{rgb}{0.72900,0.69400,0.12500}%
\begin{tikzpicture}

\begin{axis}[%
width=0.958\figurewidth,
height=\figureheight,
at={(0\figurewidth,0\figureheight)},
scale only axis,
xmin=40,
xmax=230,
xlabel style={font=\color{white!15!black}},
xlabel={$\text{D}_{\text{Deepwalk}}\text{ (or D}_{\text{PolyDeepwalk}}\text{*K)}$},
ymin=0.50,
ymax=0.59,
legend columns = 2,
grid, 
grid style={line width=.15pt, draw=white!0}, 
ylabel style={font=\color{white!15!black}},
ylabel={Micro-F1},
axis background/.style={fill=white},
legend style={legend cell align=left, align=left, draw=white!15!black, nodes={scale=0.8}, at={(1.02, 0.3)}},
axis background/.style={fill=gray!12} 
]
\addplot [color=mycolor1, mark=x, mark options={solid, mycolor1}, thick]
  table[row sep=crcr]{%
60	0.5668\\
90	0.563\\
120	0.5562\\
150	0.5561\\
180	0.556\\
210	0.5563\\
};
\addlegendentry{Deepwalk}

\addplot [color=mycolor2, mark=o, mark options={solid, mycolor2}, thick]
  table[row sep=crcr]{%
60	0.5565\\
90	0.5718\\
120	0.5775\\
150	0.5722\\
180	0.5717\\
210	0.5712\\
};
\addlegendentry{PolyDeepwalk}

\addplot [color=mycolor3, mark=asterisk, mark options={solid, mycolor3}, thick]
  table[row sep=crcr]{%
60	0.5389\\
90	0.5664\\
120	0.5719\\
150	0.5715\\
180	0.5714\\
210	0.5716\\
};
\addlegendentry{MSSG}

\addplot [color=mycolor4, mark=square, mark options={solid, mycolor4}, thick]
  table[row sep=crcr]{%
60	0.5301\\
90	0.5704\\
120	0.5694\\
150	0.5685\\
180	0.5674\\
210	0.5682\\
};
\addlegendentry{MNE}

\end{axis}
\end{tikzpicture}%
\end{subfigure}%
\begin{subfigure}[b]{0.24\textwidth}
  \setlength\figureheight{1.00in}
  \setlength\figurewidth{1.35in}
  \centering  \scriptsize
%
%
\definecolor{mycolor1}{rgb}{0.00000,0.44700,0.64100}%
\definecolor{mycolor2}{rgb}{0.65000,0.22500,0.09800}%
\definecolor{mycolor3}{rgb}{0.14700,0.59800,0.17500}%
\begin{tikzpicture}

\begin{axis}[%
width=0.951\figurewidth,
height=\figureheight,
at={(0\figurewidth,0\figureheight)},
scale only axis,
xmin=1,
xmax=13,
xlabel style={font=\color{white!15!black}},
xlabel={$\text{context size}_{\text{ }}$},
ymin=0.51,
ymax=0.61,
grid, 
grid style={line width=.15pt, draw=white!0}, 
ylabel style={font=\color{white!15!black}},
ylabel={Micro-F1},
axis background/.style={fill=white},
legend style={legend cell align=left, align=left, draw=white!15!black, nodes={scale=0.8}, at={(0.95, 0.3)}},
axis background/.style={fill=gray!12} 
]
\addplot [color=mycolor1, mark=asterisk, mark options={solid, mycolor1}, thick]
  table[row sep=crcr]{%
2	0.5553\\
4	0.558\\
6	0.5552\\
8	0.5583\\
10	0.5594\\
12	0.56\\
};
\addlegendentry{Deepwalk}

\addplot [color=mycolor2, mark=o, mark options={solid, mycolor2}, thick]
  table[row sep=crcr]{%
2	0.5767\\
4	0.58\\
6	0.574\\
8	0.5702\\
10	0.57\\
12	0.5698\\
};
\addlegendentry{PolyDeepwalk}

\end{axis}
\end{tikzpicture}%
\end{subfigure}
\vspace{-15pt}
\caption{Node classification evaluation on Flickr dataset.}
\vspace{-0pt}
\label{fig:cls_flickr}
\end{figure}
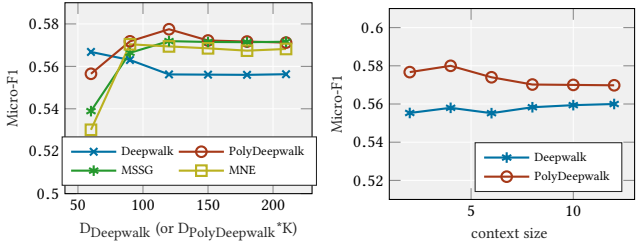

Unless specifically stated in each task, the default model hyperparameters are set as follows. For BlogCatalog dataset, the number of walks generated per node is $110$, the length of walk is $11$, the context size is $8$, the number of facets is $6$ considered for PolyDeepwalk, the embedding of dimension is $210$ for each node in Deepwalk and $210/6=35$ for each node's facet in PolyDeepwalk, the number of negative samples is $5$ for Deepwalk and $10$ for PolyDeepwalk. For Flickr dataset, the parameter settings are similar, except that the number of facets is $5$ for PolyDeepwalk, the embedding dimension is $180$ for each node in Deepwalk and $180/5=36$ for each node's facet in PolyDeepwalk, and the number of negative samples is $15$ for PolyDeepwalk. It is worth noting that the dimension of polysemous embedding is divided by the number of facets, so that after concatenation, the resultant embedding has the same length as the single-embedding counterpart. We use Support Vector Machine as the classification model, where $80\%$ of data samples are used as training data. The performance comparison is illustrated in Figure~\ref{fig:cls_bc} and Figure~\ref{fig:cls_flickr}. Note that we do not plot the performance of NMF as its F1 scores are much lower than other models. Some work includes node attributes as additional information source when using NMF related models for node classification~\cite{Huang-etal17label,Li-etal17attributed}, but this is beyond our experiment setting. From the figures, some observations are summarized as below:
\begin{itemize}[leftmargin=*]
    \item In general, PolyDeepwalk achieves better performances than other polysemous embedding models. Also, by varying the dimensionality, we can observe that polysemous embedding models are less likely to be affected, while the classification performance of Deepwalk gradually decreases as dimensionality increases.
    \item Deepwalk and PolyDeepwalk have different responses to the change of context size. The former achieves better performances as context size increases, while the latter shows the opposite trend. A possible reason is that, as the context window enlarges in PolyDeepwalk, random walks from each node are more likely to reach other network regions of different facets. As a result, the facet distributions are over smoothed, so it becomes more difficult to decide which facet a context window belongs to. It could be one of the challenges to be tackled in future work.
\end{itemize}

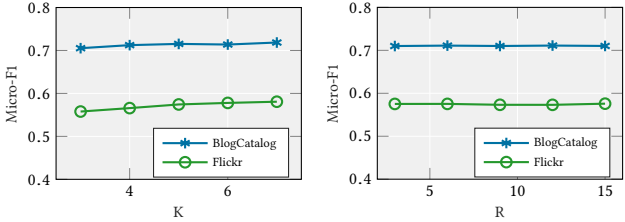
\begin{figure}[t]
\captionsetup{font=small}
\captionsetup[subfigure]{justification=centering}
\hspace{0pt}
\begin{subfigure}[b]{0.24\textwidth}
  \setlength\figureheight{0.90in}
  \setlength\figurewidth{1.35in}
  \centering  \scriptsize
%
%
\definecolor{mycolor1}{rgb}{0.00000,0.44700,0.64100}%
\definecolor{mycolor2}{rgb}{0.65000,0.22500,0.09800}%
\definecolor{mycolor3}{rgb}{0.14700,0.59800,0.17500}%
\begin{tikzpicture}

\begin{axis}[%
width=0.951\figurewidth,
height=\figureheight,
at={(0\figurewidth,0\figureheight)},
scale only axis,
xmin=2.5,
xmax=7.5,
xlabel style={font=\color{white!15!black}},
xlabel={K},
ymin=0.40,
ymax=0.80,
grid, 
grid style={line width=.15pt, draw=white!0}, 
ylabel style={font=\color{white!15!black}},
ylabel={Micro-F1},
axis background/.style={fill=white},
legend style={legend cell align=left, align=left, draw=white!15!black, nodes={scale=0.8}, at={(0.95, 0.3)}},
axis background/.style={fill=gray!12} 
]
\addplot [color=mycolor1, mark=asterisk, mark options={solid, mycolor1}, thick]
  table[row sep=crcr]{%
3	0.7053\\
4	0.7124\\
5	0.7153\\
6	0.7138\\
7	0.7185\\
};
\addlegendentry{BlogCatalog}

\addplot [color=mycolor3, mark=o, mark options={solid, mycolor3}, thick]
  table[row sep=crcr]{%
3	0.5578\\
4	0.5658\\
5	0.5743\\
6	0.5779\\
7	0.5807\\
};
\addlegendentry{Flickr}

\end{axis}
\end{tikzpicture}%
\end{subfigure}%
\begin{subfigure}[b]{0.24\textwidth}
  \setlength\figureheight{0.90in}
  \setlength\figurewidth{1.35in}
  \centering  \scriptsize
%
%
\definecolor{mycolor1}{rgb}{0.00000,0.44700,0.64100}%
\definecolor{mycolor2}{rgb}{0.65000,0.22500,0.09800}%
\definecolor{mycolor3}{rgb}{0.14700,0.59800,0.17500}%
\begin{tikzpicture}

\begin{axis}[%
width=0.951\figurewidth,
height=\figureheight,
at={(0\figurewidth,0\figureheight)},
scale only axis,
xmin=2,
xmax=16,
xlabel style={font=\color{white!15!black}},
xlabel={R},
ymin=0.40,
ymax=0.80,
grid, 
grid style={line width=.15pt, draw=white!0}, 
ylabel style={font=\color{white!15!black}},
ylabel={Micro-F1},
axis background/.style={fill=white},
legend style={legend cell align=left, align=left, draw=white!15!black, nodes={scale=0.8}, at={(0.95, 0.3)}},
axis background/.style={fill=gray!12} 
]
\addplot [color=mycolor1, mark=asterisk, mark options={solid, mycolor1}, thick]
  table[row sep=crcr]{%
3	0.7100\\
6	0.7110\\
9	0.7101\\
12	0.7111\\
15	0.7102\\
};
\addlegendentry{BlogCatalog}

\addplot [color=mycolor3, mark=o, mark options={solid, mycolor3}, thick]
  table[row sep=crcr]{%
3	0.5752\\
6	0.5753\\
9	0.5732\\
12	0.5731\\
15	0.5758\\
};
\addlegendentry{Flickr}

\end{axis}
\end{tikzpicture}%
\end{subfigure}
\vspace{-15pt}
\caption{Parameter analysis for node classification.}
\vspace{-0pt}
\label{fig:cls_self_facet}
\end{figure}

Besides making comparisons to baseline models, we also analyze the sensitivity of PolyDeepwalk to some of the key parameters when modeling node polysemy. The experiment results are shown in Figure~\ref{fig:cls_self_facet}. Some observations are made as below:
\begin{itemize}[leftmargin=*]
    \item Increasing the number of facets has positive influence on the classification performance of the proposed model. It is also worth noting that we keep the total embedding dimension (i.e., $K \times D_{\text{PolyDeepwalk}}$) to be fixed, where $D_{\text{PolyDeepwalk}}$ actually decreases as $K$ increases, so that the downstream classification task will not be affected by the feature size.
    \item In this part of experiment, changing the facet sampling rate for each observation (i.e., context window) does not significantly perturb the results. The possible reason is that many random walks have been generated from each node, so the facet sampling is also implicitly performed for each random walk. The training samples are already adequate even when we reduce the facet sampling rate for each context window.
\end{itemize}

\begin{figure}[t]
\captionsetup{font=small}
\captionsetup[subfigure]{justification=centering}
\hspace{0pt}
\begin{subfigure}[b]{0.24\textwidth}
  \setlength\figureheight{1.00in}
  \setlength\figurewidth{1.35in}
  \centering  \scriptsize
%
%
\definecolor{mycolor1}{rgb}{0.00000,0.44700,0.64100}%
\definecolor{mycolor2}{rgb}{0.65000,0.22500,0.09800}%
\definecolor{mycolor3}{rgb}{0.14700,0.59800,0.17500}%
\definecolor{mycolor4}{rgb}{0.72900,0.69400,0.12500}%
\begin{tikzpicture}

\begin{axis}[%
width=0.958\figurewidth,
height=\figureheight,
at={(0\figurewidth,0\figureheight)},
scale only axis,
xmin=40,
xmax=230,
xlabel style={font=\color{white!15!black}},
xlabel={$\text{D}_{\text{Deepwalk}}\text{ (or D}_{\text{polydeepwalk}}\text{*K)}$},
ymin=0.65,
ymax=0.74,
legend columns = 2,
grid, 
grid style={line width=.15pt, draw=white!0}, 
ylabel style={font=\color{white!15!black}},
ylabel={Micro-F1},
axis background/.style={fill=white},
legend style={legend cell align=left, align=left, draw=white!15!black, nodes={scale=0.8}, at={(1.02, 0.3)}},
axis background/.style={fill=gray!12} 
]
\addplot [color=mycolor1, mark=x, mark options={solid, mycolor1}, thick]
  table[row sep=crcr]{%
60	0.7119\\
90	0.7043\\
120	0.6987\\
150	0.6970\\
180	0.6924\\
210	0.6878\\
};
\addlegendentry{Deepwalk}

\addplot [color=mycolor2, mark=o, mark options={solid, mycolor2}, thick]
  table[row sep=crcr]{%
60	0.7186\\
90	0.7195\\
120	0.7259\\
150	0.7234\\
180	0.7226\\
210	0.7202\\
};
\addlegendentry{PolyDeepwalk}

\addplot [color=mycolor3, mark=asterisk, mark options={solid, mycolor3}, thick]
  table[row sep=crcr]{%
60	0.7061\\
90	0.7140\\
120	0.7190\\
150	0.7202\\
180	0.7198\\
210	0.7202\\
};
\addlegendentry{MSSG}

\addplot [color=mycolor4, mark=square, mark options={solid, mycolor4}, thick]
  table[row sep=crcr]{%
60	0.6847\\
90	0.7036\\
120	0.7094\\
150	0.7103\\
180	0.7122\\
210	0.7118\\
};
\addlegendentry{MNE}

\end{axis}
\end{tikzpicture}%
\end{subfigure}%
\begin{subfigure}[b]{0.24\textwidth}
  \setlength\figureheight{1.00in}
  \setlength\figurewidth{1.35in}
  \centering  \scriptsize
%
%
\definecolor{mycolor1}{rgb}{0.00000,0.44700,0.64100}%
\definecolor{mycolor2}{rgb}{0.65000,0.22500,0.09800}%
\definecolor{mycolor3}{rgb}{0.14700,0.59800,0.17500}%
\begin{tikzpicture}

\begin{axis}[%
width=0.951\figurewidth,
height=\figureheight,
at={(0\figurewidth,0\figureheight)},
scale only axis,
xmin=1,
xmax=13,
xlabel style={font=\color{white!15!black}},
xlabel={$\text{context size}_{\text{ }}$},
ymin=0.64,
ymax=0.74,
grid, 
grid style={line width=.15pt, draw=white!0}, 
ylabel style={font=\color{white!15!black}},
ylabel={Micro-F1},
axis background/.style={fill=white},
legend style={legend cell align=left, align=left, draw=white!15!black, nodes={scale=0.8}, at={(0.95, 0.3)}},
axis background/.style={fill=gray!12} 
]
\addplot [color=mycolor1, mark=asterisk, mark options={solid, mycolor1}, thick]
  table[row sep=crcr]{%
2	0.6683\\
4	0.6715\\
6	0.6784\\
8	0.6840\\
10	0.6886\\
12	0.6897\\
};
\addlegendentry{Deepwalk}

\addplot [color=mycolor2, mark=o, mark options={solid, mycolor2}, thick]
  table[row sep=crcr]{%
2	0.7168\\
4	0.7184\\
6	0.7126\\
8	0.7113\\
10	0.7074\\
12	0.7049\\
};
\addlegendentry{PolyDeepwalk}

\end{axis}
\end{tikzpicture}%
\end{subfigure}
\vspace{-15pt}
\caption{Node classification evaluation on BlogCatalog dataset.}
\vspace{-0pt}
\label{fig:cls_bc}
\end{figure}
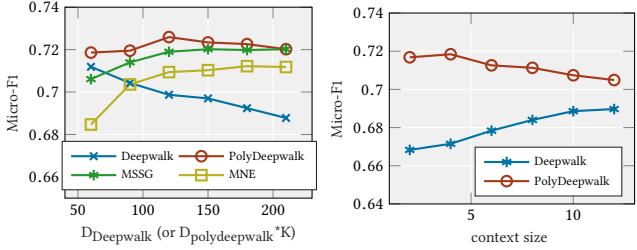

\begin{table*}[t]
\normalsize
\centering
\setlength{\tabcolsep}{2.9pt}
\ra{1.}
\begin{tabular}{@{}c|cccc|c|cccc|c@{}}\toprule
\multicolumn{1}{c}{} &
\multicolumn{5}{c}{\textbf{BlogCatalog}} & \multicolumn{5}{c}{\textbf{Flickr}} \\
\cmidrule(l){2-6} \cmidrule(l){7-11}
\multicolumn{1}{c}{} & 
\multicolumn{1}{c}{\textbf{HR@10\,\,}} & \multicolumn{1}{c}{\textbf{HR@50\,\,}} & \multicolumn{1}{c}{\textbf{HR@100}} & \multicolumn{1}{c}{\textbf{HR@200}} & \multicolumn{1}{c}{\textbf{\, AUC \,}} & \multicolumn{1}{c}{\textbf{HR@10\,\,}} & \multicolumn{1}{c}{\textbf{HR@50\,\,}} & \multicolumn{1}{c}{\textbf{HR@100}} & \multicolumn{1}{c}{\textbf{HR@200}} & \multicolumn{1}{c}{\textbf{\, AUC \,}} \\ \midrule
{\bf PolyDeepwalk} & $0.234$ & $0.493$ & $0.615$ & $0.737$ & $0.957$ & $0.210$ & $0.395$ & $0.487$ & $0.590$ & $0.928$ \\
{\bf Deepwalk} & $0.253$ & $0.512$ & $0.639$ & $0.764$ & $0.950$ & $0.195$ & $0.348$ & $0.430$ & $0.530$ & $0.912$ \\
\hline
{\bf NMF} & $0.041$ & $0.132$ & $0.196$ & $0.280$ & $0.801$ & $0.048$ & $0.148$ & $0.226$ & $0.335$ & $0.852$ \\
{\bf MSSG} & $0.102$ & $0.272$ & $0.376$ & $0.505$ & $0.910$ & $0.071$ & $0.152$ & $0.207$ & $0.268$ & $0.811$ \\
\bottomrule
\end{tabular}
\vspace{-0pt}
\caption{Performance of Homogeneous Link Prediction} \label{table:link_homo}
\vspace{-8pt}
\end{table*}

\begin{table*}[t]
\normalsize
\centering
\setlength{\tabcolsep}{2.9pt}
\ra{1.}
\begin{tabular}{@{}c|cccc|c|cccc|c@{}}\toprule
\multicolumn{1}{c}{} &
\multicolumn{5}{c}{\textbf{MovieLens}} & \multicolumn{5}{c}{\textbf{Pinterest}} \\
\cmidrule(l){2-6} \cmidrule(l){7-11}
\multicolumn{1}{c}{} & 
\multicolumn{1}{c}{\textbf{HR@10\,\,}} & \multicolumn{1}{c}{\textbf{HR@50\,\,}} & \multicolumn{1}{c}{\textbf{HR@100}} & \multicolumn{1}{c}{\textbf{HR@200}} & \multicolumn{1}{c}{\textbf{\, AUC \,}} & \multicolumn{1}{c}{\textbf{HR@10\,\,}} & \multicolumn{1}{c}{\textbf{HR@50\,\,}} & \multicolumn{1}{c}{\textbf{HR@100}} & \multicolumn{1}{c}{\textbf{HR@200}} & \multicolumn{1}{c}{\textbf{\, AUC \,}} \\ \midrule
{\bf PolyPTE} & $0.067$ & $0.210$ & $0.334$ & $0.491$ & $0.892$ & $0.062$ & $0.204$ & $0.318$ & $0.460$ & $0.919$\\
{\bf PTE} & $0.040$ & $0.143$ & $0.227$ & $0.336$ &  $0.832$ & $0.007$ & $0.030$ & $0.052$ & $0.088$ & $0.703$\\
\hline
{\bf PolyGCN} & $0.050$ & $0.158$ & $0.250$ & $0.389$ & $0.863$ & $0.036$ & $0.102$ & $0.159$ & $0.236$ & $0.801$\\
{\bf GCN} & $0.039$ & $0.130$ & $0.204$ & $0.313$ &  $0.813$ & $0.028$ & $0.059$ & $0.087$ & $0.130$ & $0.721$\\
\hline
{\bf NMF} & $0.051$ & $0.165$ & $0.265$ & $0.404$ & $0.865$ & $0.005$ & $0.021$ & $0.040$ & $0.075$ & $0.654$\\
{\bf MSSG} & $0.060$ & $0.181$ & $0.288$ & $0.427$ & $0.868$ & $0.039$ & $0.128$ & $0.199$ & $0.304$ & $0.838$ \\
\bottomrule
\end{tabular}
\vspace{-0pt}
\caption{Performance of Heterogeneous Link Prediction} \label{table:link_hetero}
\vspace{-15pt}
\end{table*}

\subsection{Link Prediction in Homogeneous Networks}
Besides node classification, we also include link prediction as one of the tasks for evaluating the embedding results. We randomly select one link for each node as test data, and use the remaining network as training data. The default settings of hyperparameters are similar to those applied in the last experiment. The main difference lies on the embedding dimensionality, where we let the embedding dimension to be $D_{\text{Deepwalk}} = D_{\text{PolyDeepwalk}}$. In other words, the embedding space of PolyDeepwalk is the same as that of Deepwalk. The default dimension value, unless otherwise stated, is chosen as $150$. The performance of link prediction is summarized in Table~\ref{table:link_homo}, from which we could observe that:
\begin{itemize}[leftmargin=*]
    \item There is no significant advantage for PolyDeepwalk on BlogCatalog dataset compared with Deepwalk. Although PolyDeepwalk achieves better performance in terms of AUC score, Deepwalk is more advantageous when evaluated using $HR@k$ and $k$ is small. However, it implicitly indicates that Deepwalk is gradually surpassed by PolyDeepwalk when $k$ increases, which means PolyDeepwalk could be better at recovering links that do not match the major connection patterns of a node.
    \item PolyDeepwalk performs better than NMF, which indicates that the effectiveness of polysemous embedding is not purely inherited from the matrix factorization step which provides prior knowledge of node facet distribution.
\end{itemize}

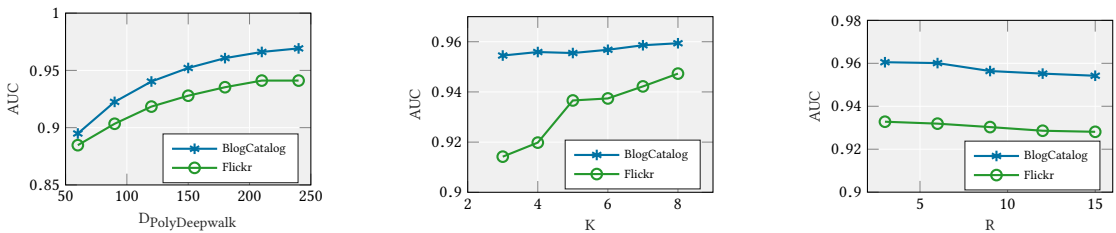
\begin{figure*}[t]
\captionsetup{font=small}
\captionsetup[subfigure]{justification=centering}
\hspace{0pt}
\begin{subfigure}[b]{0.27\textwidth}
  \setlength\figureheight{0.90in}
  \setlength\figurewidth{1.35in}
  \centering  \scriptsize
%
%
\definecolor{mycolor1}{rgb}{0.00000,0.44700,0.64100}%
\definecolor{mycolor2}{rgb}{0.65000,0.22500,0.09800}%
\definecolor{mycolor3}{rgb}{0.14700,0.59800,0.17500}%
\begin{tikzpicture}

\begin{axis}[%
width=0.951\figurewidth,
height=\figureheight,
at={(0\figurewidth,0\figureheight)},
scale only axis,
xmin=50,
xmax=250,
xlabel style={font=\color{white!15!black}},
xlabel={$\text{D}_{\text{PolyDeepwalk}}$},
ymin=0.85,
ymax=1.00,
grid, 
grid style={line width=.15pt, draw=white!0}, 
ylabel style={font=\color{white!15!black}},
ylabel={AUC},
axis background/.style={fill=white},
legend style={legend cell align=left, align=left, draw=white!15!black, nodes={scale=0.8}, at={(0.95, 0.3)}},
axis background/.style={fill=gray!12} 
]
\addplot [color=mycolor1, mark=asterisk, mark options={solid, mycolor1}, thick]
  table[row sep=crcr]{%
60	0.8949\\
90	0.9225\\
120	0.9403\\
150	0.9521\\
180	0.9607\\
210 0.9661\\
240 0.9692\\
};
\addlegendentry{BlogCatalog}

\addplot [color=mycolor3, mark=o, mark options={solid, mycolor3}, thick]
  table[row sep=crcr]{%
60	0.8847\\
90	0.9034\\
120	0.9184\\
150	0.9279\\
180	0.9353\\
210 0.9411\\
240 \\
};
\addlegendentry{Flickr}

\end{axis}
\end{tikzpicture}%
\end{subfigure}%
\hspace{10pt}
\begin{subfigure}[b]{0.27\textwidth}
  \setlength\figureheight{0.92in}
  \setlength\figurewidth{1.35in}
  \centering  \scriptsize
%
%
\definecolor{mycolor1}{rgb}{0.00000,0.44700,0.64100}%
\definecolor{mycolor2}{rgb}{0.65000,0.22500,0.09800}%
\definecolor{mycolor3}{rgb}{0.14700,0.59800,0.17500}%
\begin{tikzpicture}

\begin{axis}[%
width=0.951\figurewidth,
height=\figureheight,
at={(0\figurewidth,0\figureheight)},
scale only axis,
xmin=2,
xmax=9,
xlabel style={font=\color{white!15!black}},
xlabel={K},
ymin=0.900,
ymax=0.970,
grid, 
grid style={line width=.15pt, draw=white!0}, 
ylabel style={font=\color{white!15!black}},
ylabel={AUC},
axis background/.style={fill=white},
legend style={legend cell align=left, align=left, draw=white!15!black, nodes={scale=0.8}, at={(0.95, 0.3)}},
axis background/.style={fill=gray!12} 
]
\addplot [color=mycolor1, mark=asterisk, mark options={solid, mycolor1}, thick]
  table[row sep=crcr]{%
3	0.9545\\
4	0.9559\\
5	0.9555\\
6   0.9568\\
7	0.9586\\
8	0.9594\\
};
\addlegendentry{BlogCatalog}

\addplot [color=mycolor3, mark=o, mark options={solid, mycolor3}, thick]
  table[row sep=crcr]{%
3	0.9142\\
4	0.9198\\
5	0.9366\\
6	0.9374\\
7	0.9422\\
8   0.9473\\
};
\addlegendentry{Flickr}

\end{axis}
\end{tikzpicture}%
\end{subfigure}
\hspace{10pt}
\begin{subfigure}[b]{0.27\textwidth}
  \setlength\figureheight{0.90in}
  \setlength\figurewidth{1.35in}
  \centering  \scriptsize
%
%
\definecolor{mycolor1}{rgb}{0.00000,0.44700,0.64100}%
\definecolor{mycolor2}{rgb}{0.65000,0.22500,0.09800}%
\definecolor{mycolor3}{rgb}{0.14700,0.59800,0.17500}%
\begin{tikzpicture}

\begin{axis}[%
width=0.951\figurewidth,
height=\figureheight,
at={(0\figurewidth,0\figureheight)},
scale only axis,
xmin=2,
xmax=16,
xlabel style={font=\color{white!15!black}},
xlabel={R},
ymin=0.90,
ymax=0.98,
grid, 
grid style={line width=.15pt, draw=white!0}, 
ylabel style={font=\color{white!15!black}},
ylabel={AUC},
axis background/.style={fill=white},
legend style={legend cell align=left, align=left, draw=white!15!black, nodes={scale=0.8}, at={(0.95, 0.3)}},
axis background/.style={fill=gray!12} 
]
\addplot [color=mycolor1, mark=asterisk, mark options={solid, mycolor1}, thick]
  table[row sep=crcr]{%
3	0.9605\\
6	0.9601\\
9	0.9564\\
12	0.9552\\
15	0.9542\\
};
\addlegendentry{BlogCatalog}

\addplot [color=mycolor3, mark=o, mark options={solid, mycolor3}, thick]
  table[row sep=crcr]{%
3	0.9328\\
6	0.9319\\
9	0.9303\\
12	0.9286\\
15	0.9281\\
};
\addlegendentry{Flickr}

\end{axis}
\end{tikzpicture}%
\end{subfigure}
\vspace{-5pt}
\caption{Parameter analysis for link prediction on homogeneous networks.}
\vspace{-5pt}
\label{fig:params_lp_bc}
\end{figure*}

In addition, we also analyze the sensitivity of PolyDeepwalk by varying the values of certain hyperparameters, including dimensionality, the number of facets ($K$), and the facet sampling rate ($R$) for each context window. The results are shown in Figure~\ref{fig:params_lp_bc}. Some phenomenons are observed as below:
\begin{itemize}[leftmargin=*]
    \item Network embedding models are sensitive to embedding dimension in link prediction tasks, as we can observe a clear growth of AUC score when embedding dimension increases.
    \item Increasing the number of facets $K$ appropriately could boost link prediction performance. However, increasing $K$ will also augment the embedding storage and inference computation cost. We should better keep an appropriate $K$ value, in order to balance the performance and efficiency. 
    \item Similar to what we have discussed in node classification, increasing the facet sampling rate does not bring much benefits. Therefore, when the number of walks per node is large enough, we can keep the face sampling rate to be low in order to reduce computation costs.
\end{itemize}

\subsection{Link Prediction in Heterogeneous Networks}
In some application scenarios such as recommender systems, each link stretches across two types of nodes. Here we construct two bipartite networks from two recommendation datasets: MovieLens and Pinterest. Models such as PTE and unsupervised GCN can handle such scenario, and we extend them as PolyPTE and PolyGCN to be evaluated. For each user node, we hold out its latest interaction as the test set and use the remaining data for training.

The default values of hyperparameters, unless otherwise stated, are as follows. For both datasets, the number of facets $K$ is $5$, the embedding dimension is set as $30$, and the facet sampling rate $R$ for each observation (i.e., link) equals $K^2$. The negative sampling rate is set as $30$ for PolyPTE. Different from the last experiment where $D_{\text{PolyDeepwalk}} = D_{\text{Deepwalk}}$, the embedding dimension in this experiment is set as $D/K$. Performances of heterogeneous link prediction are shown in Table~\ref{table:link_hetero} under different metrics, where we can observe that:
\begin{itemize}[leftmargin=*]
    \item Polysemous embedding models achieves better performances than their single-embedding counterparts. It thus suggests that considering node polysemy is beneficial for modeling the association relation between different types of node entities.
    \item Polysemous embedding models in general perform better than NMF, which demonstrates that the effectiveness of polysemous models comes beyond the prior knowledge obtained from NMF.
    \item PolyGCN is not as good as PolyPTE in performance boosting. A possible reason is that we do not consider inter-facet relations in PolyGCN. However, GCN-based models still have much room for improvement, especially when attribute information is available.
\end{itemize}

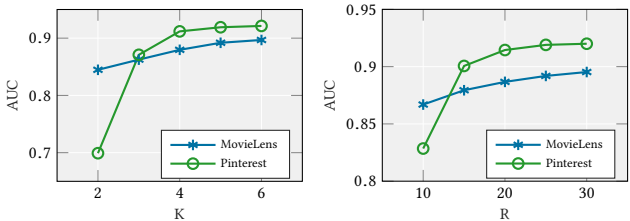
\begin{figure}[t]
\captionsetup{font=small}
\captionsetup[subfigure]{justification=centering}
\hspace{0pt}
\begin{subfigure}[b]{0.24\textwidth}
  \setlength\figureheight{0.90in}
  \setlength\figurewidth{1.35in}
  \centering  \scriptsize
%
%
\definecolor{mycolor1}{rgb}{0.00000,0.44700,0.64100}%
\definecolor{mycolor2}{rgb}{0.65000,0.22500,0.09800}%
\definecolor{mycolor3}{rgb}{0.14700,0.59800,0.17500}%
\begin{tikzpicture}

\begin{axis}[%
width=0.951\figurewidth,
height=\figureheight,
at={(0\figurewidth,0\figureheight)},
scale only axis,
xmin=1,
xmax=7,
xlabel style={font=\color{white!15!black}},
xlabel={K},
ymin=0.65,
ymax=0.95,
grid, 
grid style={line width=.15pt, draw=white!0}, 
ylabel style={font=\color{white!15!black}},
ylabel={AUC},
axis background/.style={fill=white},
legend style={legend cell align=left, align=left, draw=white!15!black, nodes={scale=0.8}, at={(0.95, 0.3)}},
axis background/.style={fill=gray!12} 
]
\addplot [color=mycolor1, mark=asterisk, mark options={solid, mycolor1}, thick]
  table[row sep=crcr]{%
2	0.8447\\
3	0.8625\\
4	0.8796\\
5	0.8919\\
6	0.8968\\
};
\addlegendentry{MovieLens}

\addplot [color=mycolor3, mark=o, mark options={solid, mycolor3}, thick]
  table[row sep=crcr]{%
2	0.6991\\
3	0.8709\\
4	0.9118\\
5	0.9190\\
6	0.9213\\
};
\addlegendentry{Pinterest}

\end{axis}
\end{tikzpicture}%
\end{subfigure}%
\begin{subfigure}[b]{0.24\textwidth}
  \setlength\figureheight{0.90in}
  \setlength\figurewidth{1.35in}
  \centering  \scriptsize
%
%
\definecolor{mycolor1}{rgb}{0.00000,0.44700,0.64100}%
\definecolor{mycolor2}{rgb}{0.65000,0.22500,0.09800}%
\definecolor{mycolor3}{rgb}{0.14700,0.59800,0.17500}%
\begin{tikzpicture}

\begin{axis}[%
width=0.951\figurewidth,
height=\figureheight,
at={(0\figurewidth,0\figureheight)},
scale only axis,
xmin=5,
xmax=35,
xlabel style={font=\color{white!15!black}},
xlabel={R},
ymin=0.80,
ymax=0.95,
grid, 
grid style={line width=.15pt, draw=white!0}, 
ylabel style={font=\color{white!15!black}},
ylabel={AUC},
axis background/.style={fill=white},
legend style={legend cell align=left, align=left, draw=white!15!black, nodes={scale=0.8}, at={(0.95, 0.3)}},
axis background/.style={fill=gray!12} 
]
\addplot [color=mycolor1, mark=asterisk, mark options={solid, mycolor1}, thick]
  table[row sep=crcr]{%
10	0.8669\\
15	0.8794\\
20	0.8867\\
25	0.8919\\
30	0.8953\\
};
\addlegendentry{MovieLens}

\addplot [color=mycolor3, mark=o, mark options={solid, mycolor3}, thick]
  table[row sep=crcr]{%
10	0.8285\\
15	0.9006\\
20	0.9145\\
25	0.9190\\
30	0.9200\\
};
\addlegendentry{Pinterest}

\end{axis}
\end{tikzpicture}%
\end{subfigure}
\vspace{-15pt}
\caption{Parameter analysis for PolyPTE.}
\vspace{-0pt}
\label{fig:link_movielens}
\end{figure}

Similar to previous experiment tasks, here we also analyze the sensitivity of PolyPTE to model hyperparameters, as shown in Figure~\ref{fig:link_movielens} for MovieLens and Pinterest. Besides some similar observations as in the last experiment, PolyPTE is sensitive to the facet sampling rate, especially when its value is low. A possible reason for such contrasting observations between PolyDeepwalk and PolyPTE could be the different forms of observation samples $o \in \mathcal{O}$. PolyDeepwalk samples plenty of random walks from each node (i.e., $o = \{\mathcal{N}(v), v\}$), while PolyPTE uses edge based sampling strategy (i.e., $o = (v_i, v_j)$). Therefore, if the facet sampling rate is low, PolyPTE could not fully consider all the possible correlation patterns between different facets of users and items. It will thus negatively affect the embedding performance.

\section{Related Work}
Network embedding models have received a lot of attention recently due to its effectiveness in learning representations for nodes in network data. Existing work on network embedding can be categorized based on various criteria. Many basic models such as Deepwalk~\cite{Perozzi-etal14deepwalk}, LINE~\cite{Tang-etal15line}, node2vec~\cite{Grover-Leskovec17node2vec} and SDNE~\cite{Wang-etal16SDNE} focus on analyzing plain networks, while more recent work starts tackling more complex networks such as considering attributes~\cite{Huang-etal17label, Liao-etal18attributed}, community structures~\cite{Wang-etal17communitypreserve} and node heterogeneity~\cite{Dong-etal17metapath2vec, Tang-etal15PTE}. Many methods solve the embedding problem by resorting to matrix factorization~\cite{Qiu-etal18unifying}, either explicitly~\cite{Ou-etal16asymmetric} or implicitly~\cite{Perozzi-etal14deepwalk, Tang-etal15line}, while recently feed-forward modules such as graph convolutional networks~\cite{Hamilton-etal17inductive,Kipf-Welling17semisupervised} are attracting more and more attention. In addition, some challenges that have been tackled in network embedding include model scalability improvement~\cite{Ying-etal18graph}, modeling dynamic networks~\cite{Zhou-etal18dynamic}, considering human-in-the-loop scenarios~\cite{Huang-etal18exploring}. Finally, network embedding has been applied in applications such as recommender systems~\cite{Grbovic-etal18realtime}, fraud detection~\cite{Wang-etal18deep} and behavioral modeling~\cite{Wang-etal18multitype}.

Some recently proposed models, not limited to network analysis, start to utilize the structural knowledge for representation learning. Ma et al. make use of hierarchical taxonomy to regularize node representation learning~\cite{Ma-etal18hierarchical}, which does not solve the same problem as in this paper since each node is still assigned only one embedding vector. Yang et al. extends matrix factorization models to explore multiple facets of homogeneous networks towards embedding~\cite{Yang-etal18multifacet}, where the method may not be extended to other base models or scenarios. Some other work such as~\cite{Zhang-etal18taxogen} and~\cite{Liu-etal18interpretation} try to extract hierarchical taxonomy from embedding results, which also do not solve the same problem as in this paper.

The property of multiple facets owned by network nodes can be analogized to a similar phenomenon of word polysemy in natural languages~\cite{Reisinger_Mooney10multiPrototype}. However, the modeling methods cannot be directly applied to our problem of network analytics due to various incompatibility issues~\cite{Liu-etal15topical, Huang-etal12multipleEmbedding}, such as the differences in language models and network models, the difference of sample formats, and the absence of the concept of documents in network data. Also,~\cite{Neel-etal15efficient} does not consider polysemy in the context vectors or the correlation between different word senses. Some relevant problems have been discussed in Heterogeneous Information Networks~\cite{Shi-etal18easing}, but it is hard to adapt the methodology to other scenarios.

\section{Conclusion and Future Work}\label{sec:conclusion}
In this paper, we propose a polysemous embedding method for considering multiple facets of nodes in network embedding. After estimating the degrees of association between each node and its facets, each facet of a node owns a embedding vector. The proposed method is flexible as it can be used to extend various existing embedding models. During the training process, embedding vectors of different facets are updated, based on the observation where the nodes locate. We also introduce how to combine multiple embedding vectors of each node during classification and link prediction. Experiments on real-world datasets comprehensively evaluate when we benefit from considering node polysemy compared with single-vector embedding baseline methods. Some possible future work are as follow: (1) Exploring the possibility of doing node facets association and polysemous embedding simultaneously, (2) Applying polysemous network embedding models to more complicated network systems such as knowledge graphs, (3) Developing optimization algorithms to further accelerate the training procedure. 

\section*{Acknowledgments}
The work is, in part, supported by NSF ({\#}IIS-1657196, {\#}IIS-1718840, {\#}IIS-1750074). The views and conclusions contained in this paper are those of the authors and should not be interpreted as representing any funding agencies.

\bibliographystyle{named}
\bibliography{KDD19_ninghao}

\end{document}